\documentclass[twocolumn]{aastex62}

\usepackage{color}

\shorttitle{Na spread in NGC\,188}
\shortauthors{A.E. Piatti}

\begin{document}

\def\msun{\hbox{M$_\odot$}}
\def\noao{The NOAO Deep Wide Field Survey MOSAIC Data Reductions, 
http://www.noao.edu/noao/noaodeep/ReductionOpt/frames.html}%
\def\submitted{AJ, submitted.}

\title{On the APOGEE DR14 sodium spread in the Galactic open cluster NGC\,188}

\author[0000-0002-8679-0589]{Andr\'es E. Piatti}
\affiliation{Consejo Nacional de Investigaciones Cient\'{\i}ficas y T\'ecnicas, Godoy Cruz 2290, C1425FQB, 
Buenos Aires, Argentina}
\affiliation{Observatorio Astron\'omico de C\'ordoba, Laprida 854, 5000, 
C\'ordoba, Argentina}
\correspondingauthor{Andr\'es E. Piatti}
\email{e-mail: andres.piatti@unc.edu.ar}

\begin{abstract}
Since several years ago, the search for multiple populations (MPs) in Galactic
open clusters (OCs) has become a field of increasing interest, particularly to
the light of the general knowledge that MPs are observed in relatively massive
clusters ($\ge$ 5 10$^4$ $M_\odot$). I report here measurements
of stellar parameters and abundances of 17 
different chemical species, including
light and iron-peak elements of the old open cluster NGC\,188, from the APOGEE 
DR14 database. I selected 15
bonafide cluster red giants that reveal an spread in [Na/Fe] of
$\pm$0.16 dex, caused by two different
groups of measurements whose mean [Na/Fe] values differ by $\sim$ 0.30 dex. A hint for a subtle
anti-correlation with Al and Ti\,II, which also show relative large [X/Fe]
dispersions, is also seen in the data. 
However,  [Na/Fe] abundances have been found
to be typically affected by uncertainties of 0.16 dex.
Furthermore, I directly compared the available spectra for stars with similar 
atmospheric parameters and it would seem that they look similar.
Therefore, I warn users of large spectroscopic surveys to be 
extra careful when finding peculiar abundance results. 
Analysis pipelines that run in an unsupervised fashion may produce bad results which may 
not be noticed before publication.
\end{abstract}


\section{Introduction}

Multiple populations (MPs) is a phenomenon commonly observed among Galactic
globular  clusters (GCs), which is characterized by the existence of stellar
populations with distinctive chemical  abundance patterns 
\citep{grattonetal2004,bl2018,marinoetal2019}. Some few
studied much younger clusters have also shown chemically different stellar
populations, which extends the MP phenomenon  down to ages of $\sim$ 2 Gyr
\citep{pancino2018,martocchiaetal2018a,hollyheadetal2019,lianddegrij2019}.

As far as Galactic open clusters (OCs) are considered, some searches for MPs
have been  attempted \citep[see e.g.][]{dasilvaetal2009,pancinoetal2010,carreraetal2013}.
\citet{bragagliaetal2012} studied spectroscopically the 6 Gyr old OC
Berkeley\,39  and concluded that it is a single-population OC. Other 1.0-2.6 Gyr
OCs (NGC\,2158, 2420, 2682 and Berkeley\,29) have also been studied by
\citet{carreraandmartinez2013}, who did not detect any anomalous spreads nor
bimodalities in their CN and CH distributions.  More recently,
\citet{cantatgaudinetal2014} studied the nearly 300 Myr old OC NGC\,6705 as part
of the {\it Gaia}-ESO survey, and found no sign of the expected MP behaviors
between Al, Mg, Si, and Na abundances. 

Here I analyze  evidence of intrinsic spreads in the abundance
of sodium -- and to a lesser extent of Al and Si -- in the old
OC NGC\,188 \citep[age = 6.3$\pm$0.2 Gyr,][]{bonatto2019}.  This
finding, would suggest
a large age range to the lower limit of appearance of MPs, perhaps modulated by
other parameters such as mass and metallicity.

This work is organized as follows. Section 2 describes the observational material 
employed and the criteria used to select a {\it bonafide} sample of red giant members.
In Section 3 I analyze the selected data and discuss the observed hints of MPs
in NGC\,188. Finally, Section 4 summarises the main conclusions of this work.

\section{Collected Data }

I made use of stellar parameters and element abundances for 19 chemical species
publicly available as part of the Sloan Digital Sky Survey IV, 
in particular of APOGEE DR14
\citep{blantonetal2017,abolfathietal2018}.
The aforementioned stellar properties
have been determined by the APOGEE Stellar Parameters and Abundances  pipeline
\citep[ASPCAP,][]{garciaperezetal2016} from high-resolution ($R$ = 22.500)
near-infrared ($\lambda$$\lambda$1.51-1.69$\mu$m) spectra and downloaded from 
the APOGEE webpage\footnote{http://www.sdss.org/dr14/}. 
Observed spectra are used by ASPCAP to determine atmospheric parameters and chemical abundances by comparing them to libraries of theoretical spectra, using 
$\chi^2$ minimization in a multidimensional parameter space. Firstly, stellar 
parameters are derived from the entire APOGEE spectral range, and then the determination
of individual chemical abundances  is done from spectral windows
optimized for each element.

I searched for any entry in the \texttt{allStar-l31c.2} catalogue with the \texttt{FIELD}
column containing the string \texttt{N188}. As a result, I gathered 758 stars distributed 
within a radius of $\sim$ 1.6$\degr$ from the cluster center and with 2MASS $K_s$ mag down to
$\sim$ 12.5 mag. For each star I collected its  radial velocity (RV), effective temperature 
($T_{eff}$), surface gravity (log($g$)) and the chemical abundance ratios  [C/Fe], [N/Fe],
[O/Fe], [Na/Fe], [Mg/Fe], [Al/Fe], [Si/Fe], [P/Fe], [S/Fe], [K/Fe], [Ca/Fe], [Ti II/Fe], [V/Fe],
[Cr/Fe], [Mn/Fe], [Co/Fe], [Ni/Fe] and [Fe/H], with their respective uncertainties.

\begin{figure}
     \includegraphics[width=\columnwidth]{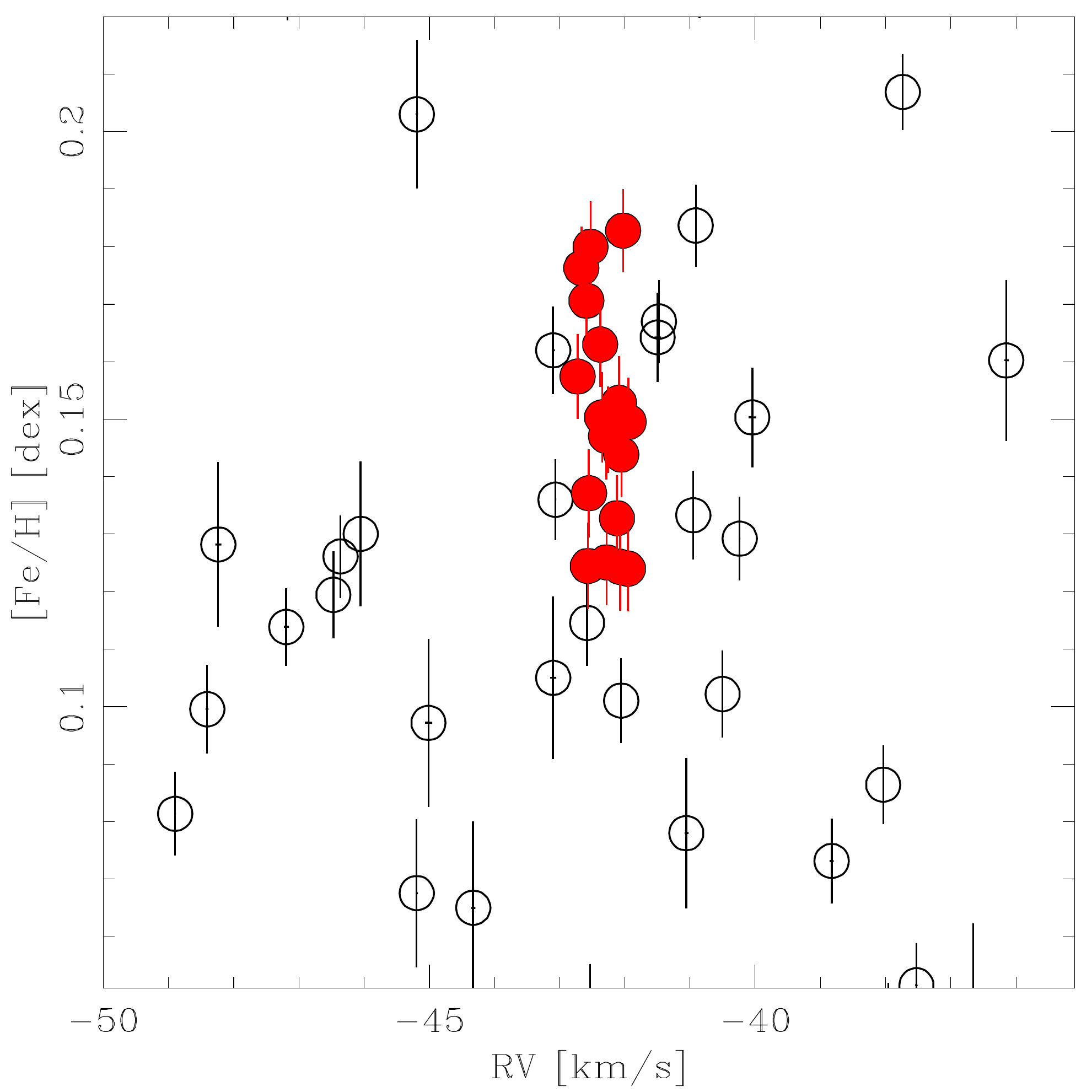}
\caption{RV versus [Fe/H] diagnostic diagram. Error bars are also drawn. Red filled
circles represent the selected cluster members.}
 \label{fig1}
\end{figure}

In order to disentangle {\it bonafide} cluster members from field stars I used
the diagnostic  plane [Fe/H] versus RV (see Fig.~\ref{fig1}), which clearly
highlights a group of stars  tightly distributed around the known cluster RV =
42.4$\pm$0.1 km/s \citep{chumaketal2010}  and metal content
0.12 $<$ [Fe/H] (dex) $<$ 0.20 \citep{hillsetal2015,bonatto2019}). The resulting mean
and dispersion of the 15 selected stars (red filled circles,  Table~\ref{tab:table1}) turned out to be
$<$RV$>$ = -42.30 $\pm$ 0.048 km/s, $\sigma$$_{\rm RV}$ = 0.23 $\pm$ 0.05
km/s,   $<$[Fe/H]$>$ =  +0.16 $\pm$ 0.01 dex and $\sigma$$_{\rm [Fe/H]}$ = 0.02
$\pm$ 0.01 dex, respectively. They are red giant stars with 4100 $\la$ $T_{eff}$
(K) $\la$ 5000  and  1.5 $\la$ log($g$) $\la$ 3.5, distributed  within a circle
with a radius of 25$\arcmin$, which is nearly half of the cluster tidal radius
\citep[44.78$\arcmin$,][]{wangetal2015}. 

A search into the {\em Gaia} DR2 archive \citep{gaiaetal2016,gaiaetal2018b} provided for the  15 {\em bonafide}
members: $<\mu_{\rm{RA}}>=-2.32\pm0.11$~mas/yr (median=--2.36);
$<\mu_{\rm{Dec}}>=-0.90\pm0.12$~mas/yr (median=--0.91); and
$<\varpi>=0.51\pm0.03$~mas (median=0.51). The mean proper motions and parallax
are fully compatible with the ones found in the literature 
\citep{plataiseal2003}.  I found that there are no apparent subgroups or bi-modalities in the data.

\begin{deluxetable}{lcc}
\tablecaption{APOGEE IDs and coordinates of the selected stars.  \label{tab:table1}}
\tablehead{
\colhead{ID} & \colhead{R.A. ($\degr$)} & \colhead{Dec. (\degr) }
}
\startdata
 2M00320079+8511465 &  8.0032949 &85.1962433\\
 2M00415197+8527070 &10.4665461 &85.4519730\\
 2M00444460+8532163 &11.1858711 &85.5378723\\
 2M00454489+8504180 &11.4370728 &85.0716934\\
 2M00463920+8523336 &11.6633635 &85.3926849\\
 2M00533497+8511145 &13.3957109 &85.1873856\\
 2M00533572+8520583 &13.3988371 &85.3495483\\
 2M00541152+8515231 &13.5480270 &85.2564316\\
 2M00542287+8455398 &13.5953159 &84.9277420\\
 2M00543664+8501152 &13.6526823 &85.0208817\\
 2M00571844+8510288 &14.3268414 &85.1746750\\
 2M00581691+8540183 &14.5704737 &85.6717606\\
 2M01003483+8503198 &15.1451283 &85.0555344\\
 2M01015206+8506329 &15.4669333 &85.1091614\\
 2M01025280+8517563 &15.7200136 &85.2989655\\
\enddata
\end{deluxetable}

\section{Analysis and discussion}

\begin{figure*}
     \includegraphics[width=11.cm]{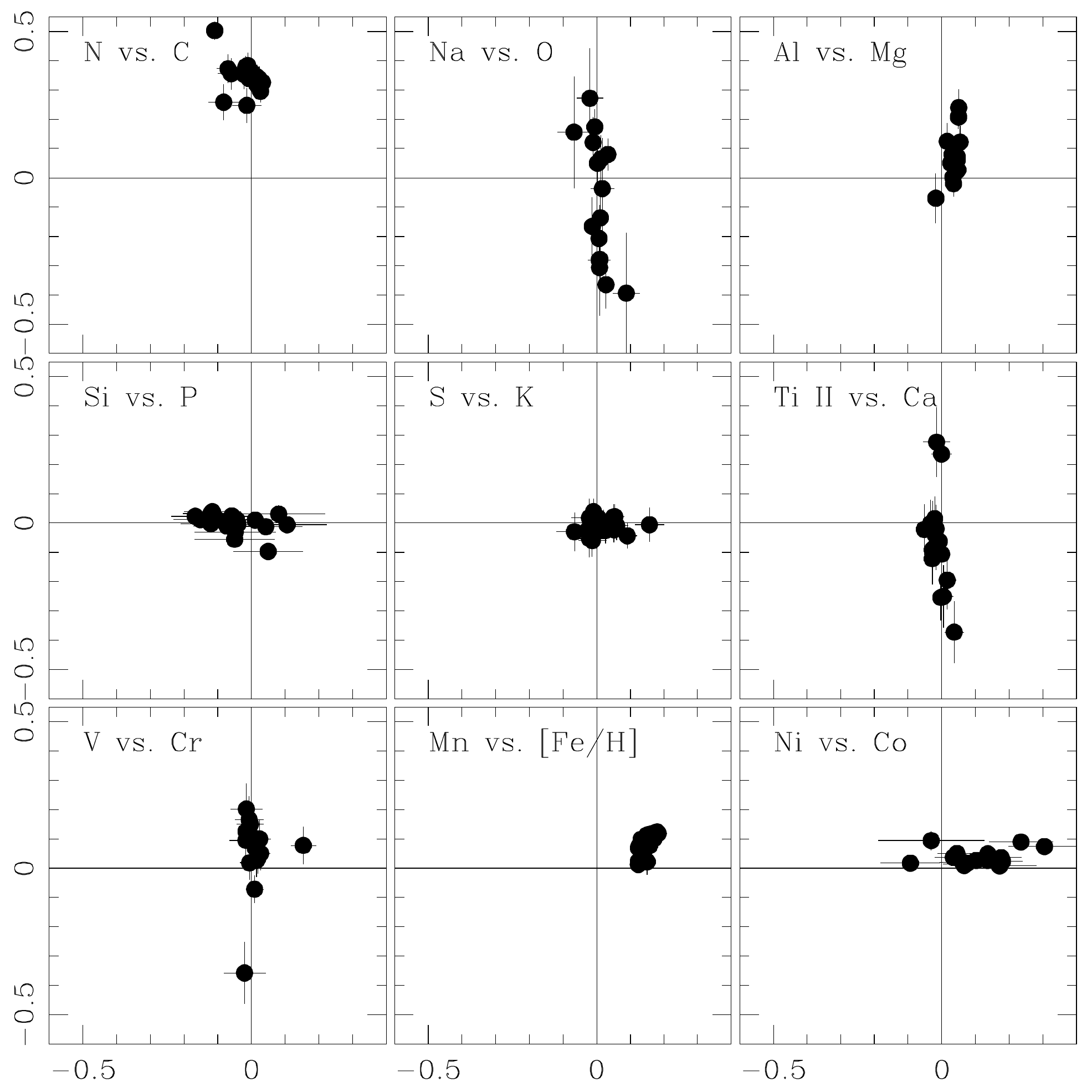}
      \includegraphics[width=4.075 cm]{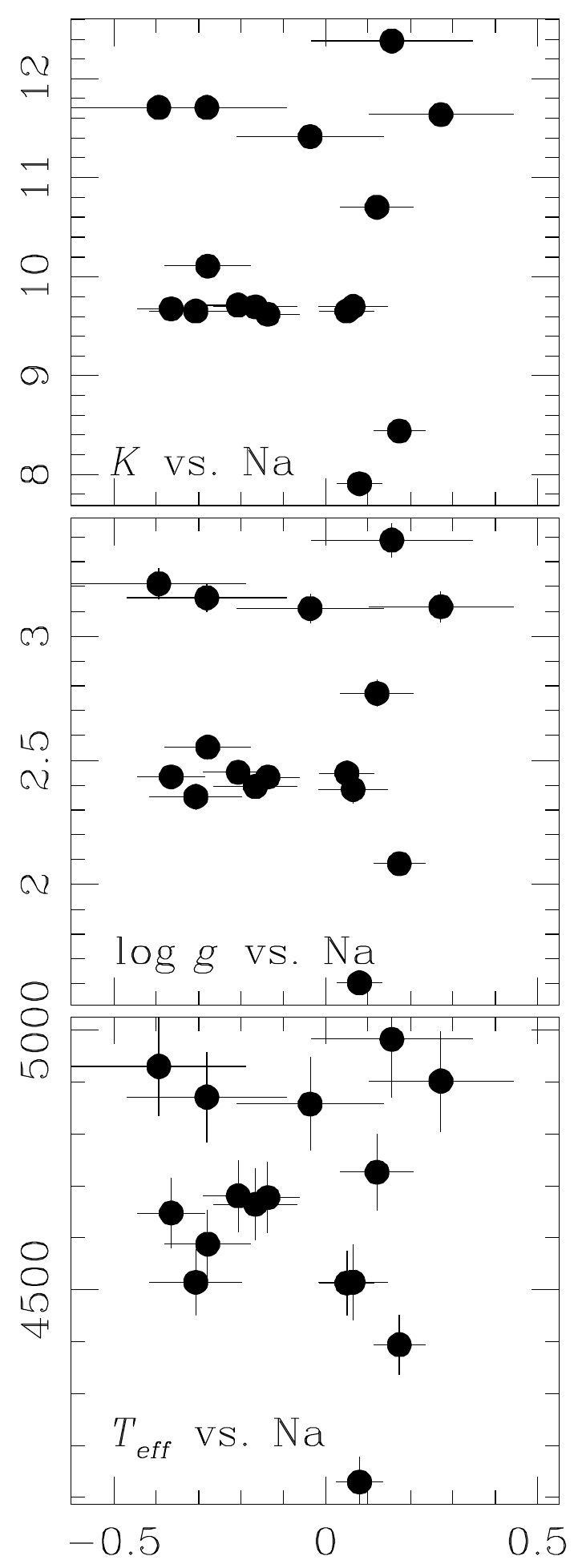}
\caption{ {\it Left:} Relationships between different [X/Fe] ratios. The chemical 
(X) species are indicated at the top-left margin of each panel. Error bars are also drawn. {\it Right:} 2MASS $K$ mag (mag), log($g$) and $T_{eff}$
(\degr K)  versus [Na/Fe] are drawn from top to bottom panels, respectively.}
 \label{fig2}
\end{figure*}

Fig.~\ref{fig2} shows the relationships between [X/Fe] ratios for the
different  chemical species (X) considered. At first glance, it seems that they
span different ranges, from a negligible  dispersion (e.g., 0, Mg, S,
P) up to visible spreads (e.g., Na, Ti\,II, V, Co). A similarly
uneven behavior, especially in the light elements C, N, O, Na, Mg, and Al, is
the fingerprint of MPs in GCs: [Na/Fe] is the flagship among these elements as a
good indicator of the existence of MPs, with spreads of up to 0.6~dex and more
and with no GC having a null spread. Conversely, O, Mg, Si, among others, do not
necessary show a spread to make evident the existence of MPs \citep{
carrettaetal2009,meszarosetal2015,pancinoetal2017},
especially for metal-rich clusters. Another flagship element is [N/Fe], because
it has never a negligible spread in GCs \citep{pancinoetal2010,meszarosetal2015}, 
but here I do
not see any significant variation, which is puzzling and requires further
investigation.

In order to quantify the  spreads of our [X/Fe]
ratios, I used the maximum likelihood approach by optimising the probability
$\mathcal{L}$ that  the sample of selected stars with [X/Fe]$_i$ ratios  and
errors $\sigma_i$ are  drawn from a population with mean $<$[X/Fe]$>$ and
dispersion W   \citep[e.g.,][]{pm1993,walker2006}, as follows:\\

$\mathcal{L}\,=\,\prod_{i=1}^N\,\left( \, 2\pi\,(\sigma_i^2 + W^2 \, ) 
\right)^{-\frac{1}{2}}\,\exp \left(-\frac{({\rm [X/Fe]_i} \,- <{\rm [X/Fe]}>)^2}{2(\sigma_i^2 + W^2)}
\right)$\\

\noindent where the errors on the mean and dispersion were computed from the
respective covariance matrices. Table~\ref{tab:table2} lists the resulting mean
$<$[X/Fe]$>$ ratios and the respective  spreads W. As can be seen, the
dispersion derived for Na, and also for Al and  Ti\,II, reveal that there exist
 spreads. Additionally, some iron-peak elements (V, Co) also resulted
with clear spreads in their mean $<$[X/Fe]$>$ ratios. In the
case of titanium, I note that Ti~I does not show any spread
 ($<$[Ti\,I/Fe]$>$= 0.02$\pm$0.01 dex, W = 0.00$\pm$0.02 dex), thus 
casting some
doubt on the robustness of the Ti~II measurements, also in light of a slight
trend observed with T$_{eff}$. Similar trends are also observed for V and Co,
while a flat trend is observed for Na and Al with both T$_{eff}$ and log$g$.

\begin{figure*}
     \includegraphics[width=0.9\columnwidth]{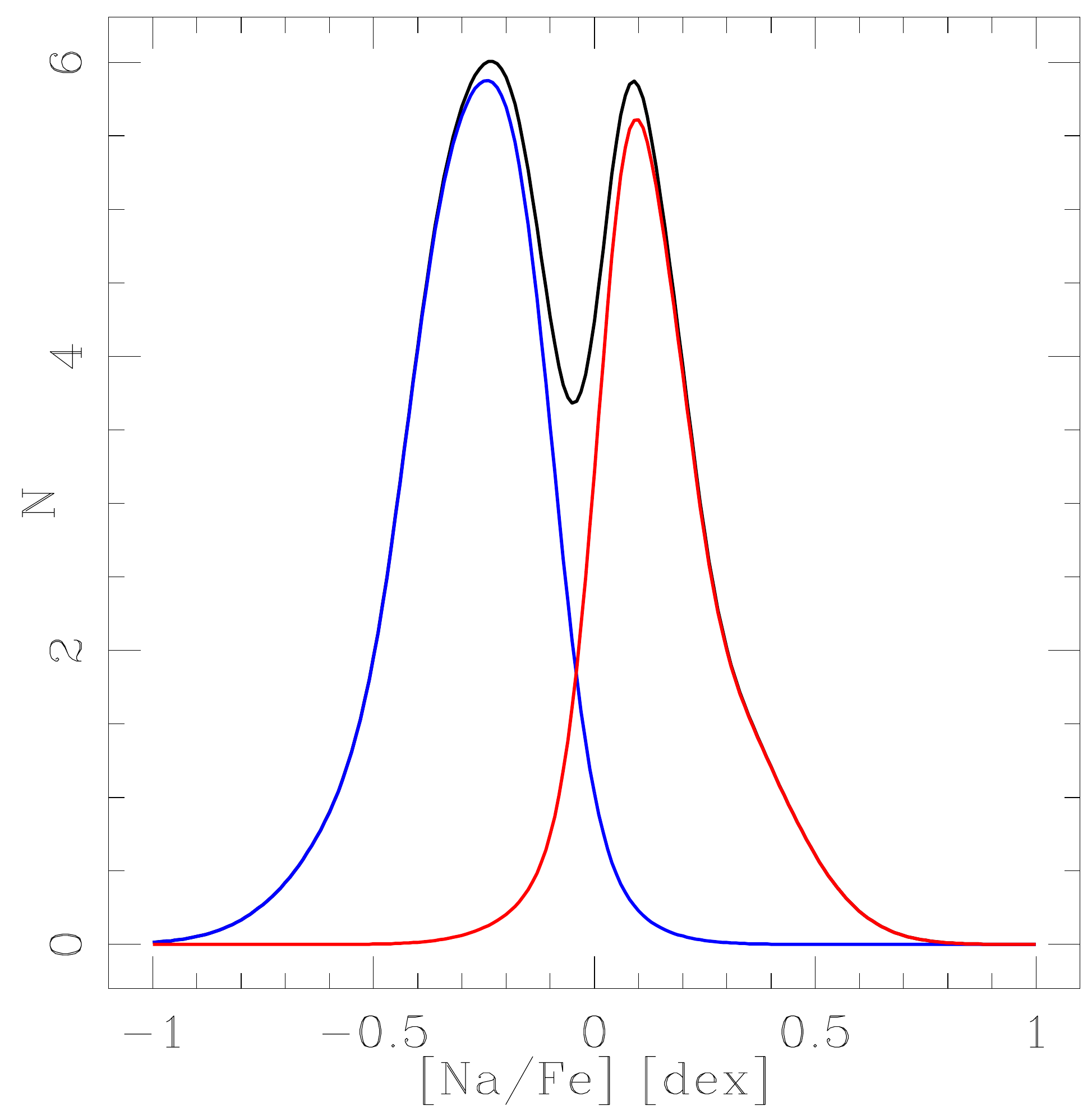}
      \includegraphics[width=0.9\columnwidth]{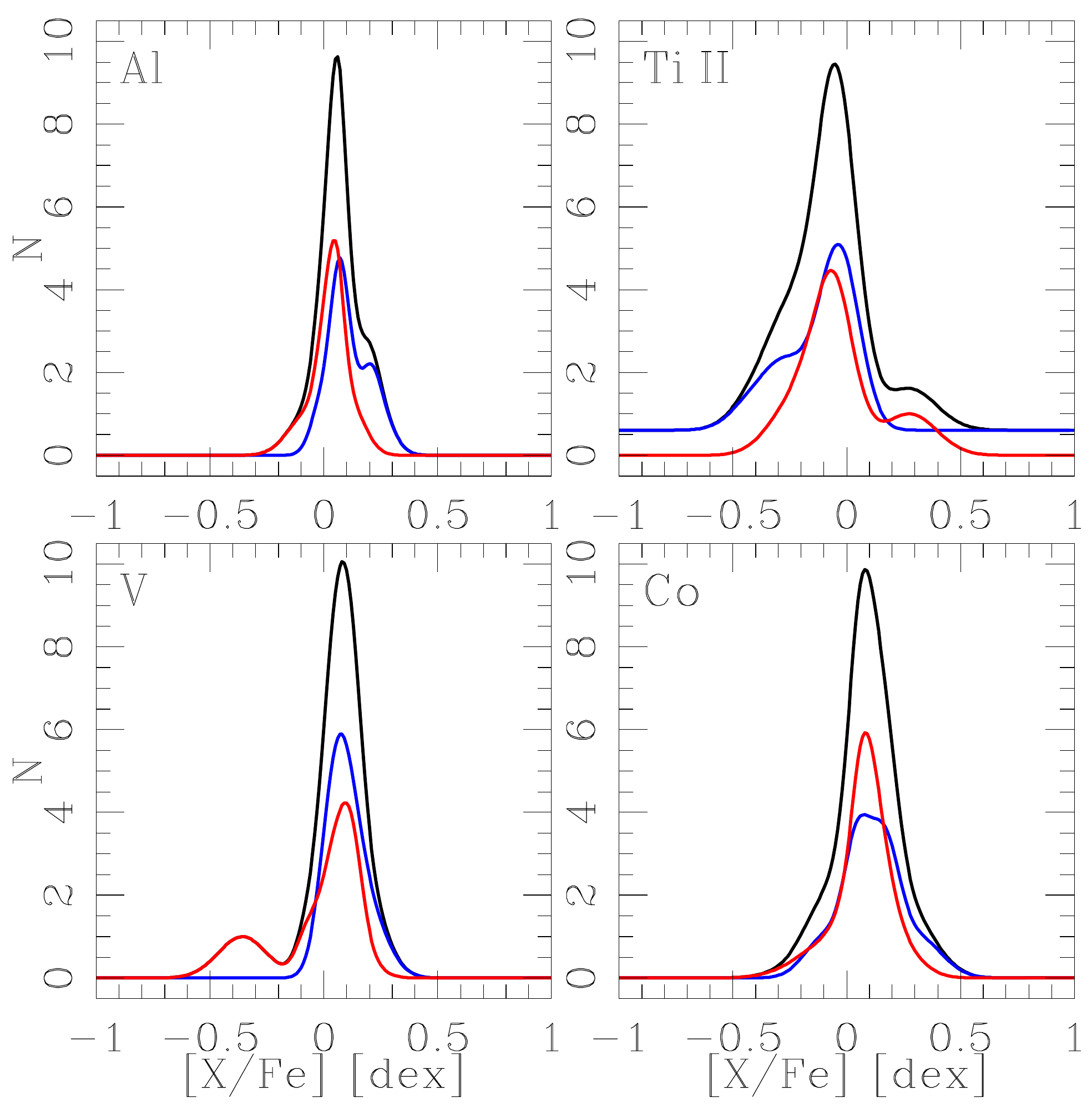}
\caption{[X/Fe] distributions for different chemical species. Blue and red lines
correspond to Na-poor and Na-rich stars (see text for details), respectively; black lines being
the sum of them.}
 \label{fig3}
\end{figure*}

I then built [X/Fe] distributions for Na, Al, Ti\,II, V and Co by summing all
the individual [X/Fe] values. I represented each [X/Fe] value by a Gaussian
function with center and full-width half maximum equal to the [X/Fe] value and 
2.355 times the associated error, respectively, and assigned to each Gaussian
the same mean intensity. Thus, I avoid
the problem of the histogram dependence on the bin size and the end points of
bin, which frequently lead to a difficulty in interpreting the
results. Furthermore, I here take into  account the individual [X/Fe]
uncertainties. The result is a continuous distribution -- instead of a
discrete histogram -- that allows to appreciate the finest
structures. Fig.~\ref{fig3} depicts the resulting [X/Fe] distributions. As
can be seen, the [Na/Fe] dispersion derived from the maximum likelihood
statistics is indeed bimodal, with two prominent peaks differing
by $\simeq$0.3 dex. I adopted the value [Na/Fe] = -0.05 dex  as the boundary to
distinguish them, which I will call Na-poor and Na-rich stars, respectively. 

 I performed additional checks in order to discard any artifacts in the data
as responsible of the observed bimodality. I also searched all
the ASPCAP flags and the {\em Gaia} astrometric flags for peculiarities in our
sample stars, and could not find any. The spectra of Na-poor and Na-rich stars
span indistinguishably SNR ranges $\approx$ 100-600, they have \texttt{ASPCAPFLAG} and \texttt{NA$\_$FE$\_$FLAG} values equals to zero, 
which means that there were no
issues in the determination of the stellar parameters, and mean $\chi^2$ from
the ASPCAP fit $\approx$ 8.0. The [Na/Fe] abundances have been derived from two
Na features at about 16378 and 16393 \AA, respectively  \citep[see Table 3 in][]{garciaperezetal2016}.

Na-poor and Na-rich stars seem to show a suggestive
anti-correlation with a possibly bimodal Al distribution  (Pearson
correlation coefficient = -0.5), while the situation
for Ti~II, V, and Co is less clear. The derived W$_{\rm V}$ and W$_{\rm Co}$
values (see Table~\ref{tab:table2}) appear to come rather from the presence of
one star  -- a different one for V and Co -- that statistically  makes wider the
[X/Fe] ranges. I discard these chemical species as showing intrinsic  spreads
due to the small number statistics. 

\citet{carrettaetal10} showed that there is a minimum (present-day) threshold
mass of about a few 10$^4$ $M_\odot$, for the MP phenomenon to come to the
light. Indeed, as far as I am aware, NGC\,6535 is the least massive GCs with evidence of MPs. Its estimated mass is $\sim$ 3.4 10$^4$ $M_\odot$
\citep{bragagliaetal2017}. In the Galactic open cluster system, Berkeley\,39 was
targeted by \citet{bragagliaetal2012} to explore the robustness of such a
minimum mass (see their figure 1). Berkeley\,39 is as old as NGC\,188 (6 Gyr),
with  an estimated mass of $\sim$ 2 10$^4$ $M_\odot$. Finally, they did not
detect any trace of MPs, confirming \citet{carrettaetal10}'s
results.

\begin{figure}
     \includegraphics[width=0.9\columnwidth]{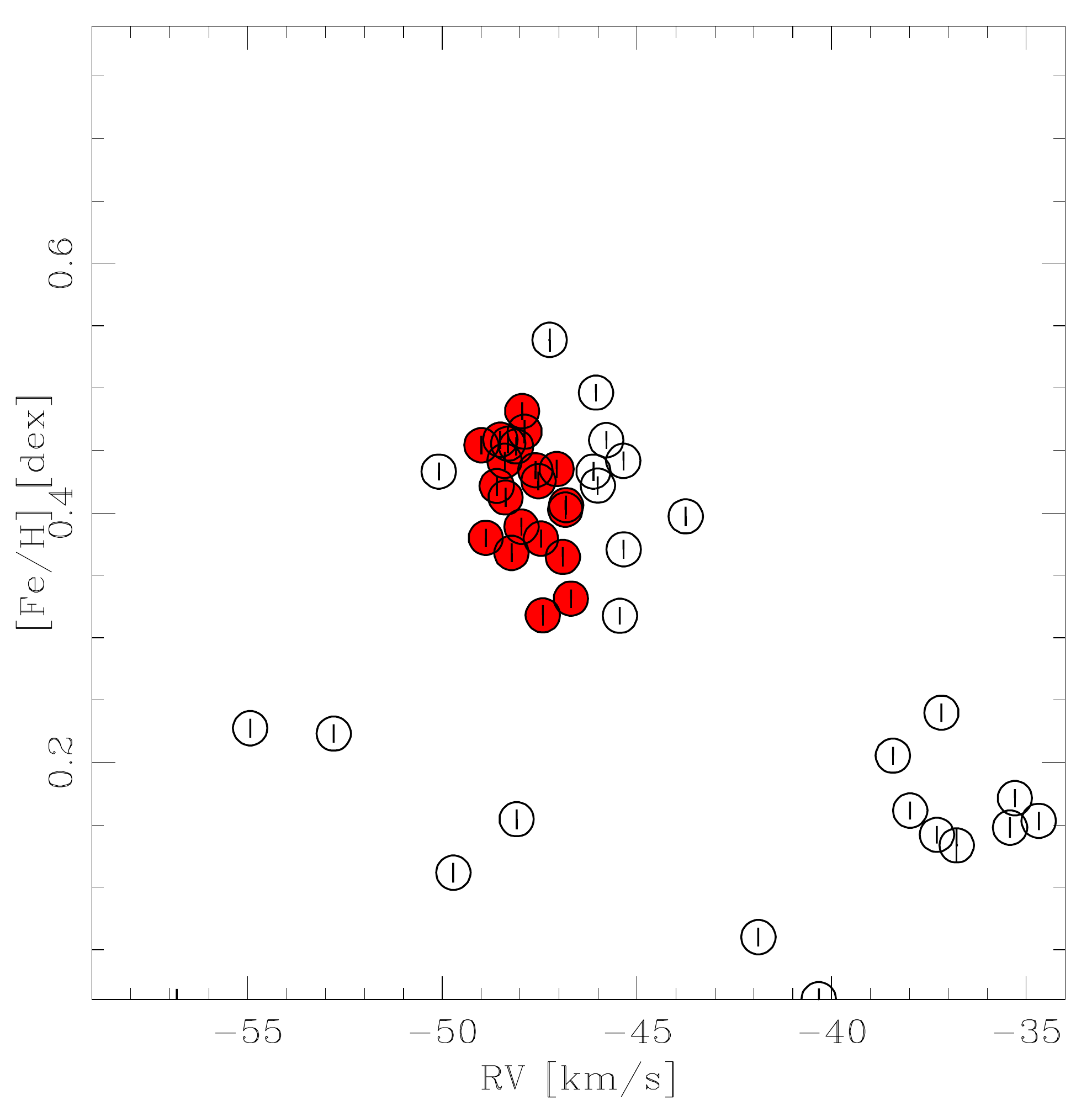}
      \includegraphics[width=0.88\columnwidth]{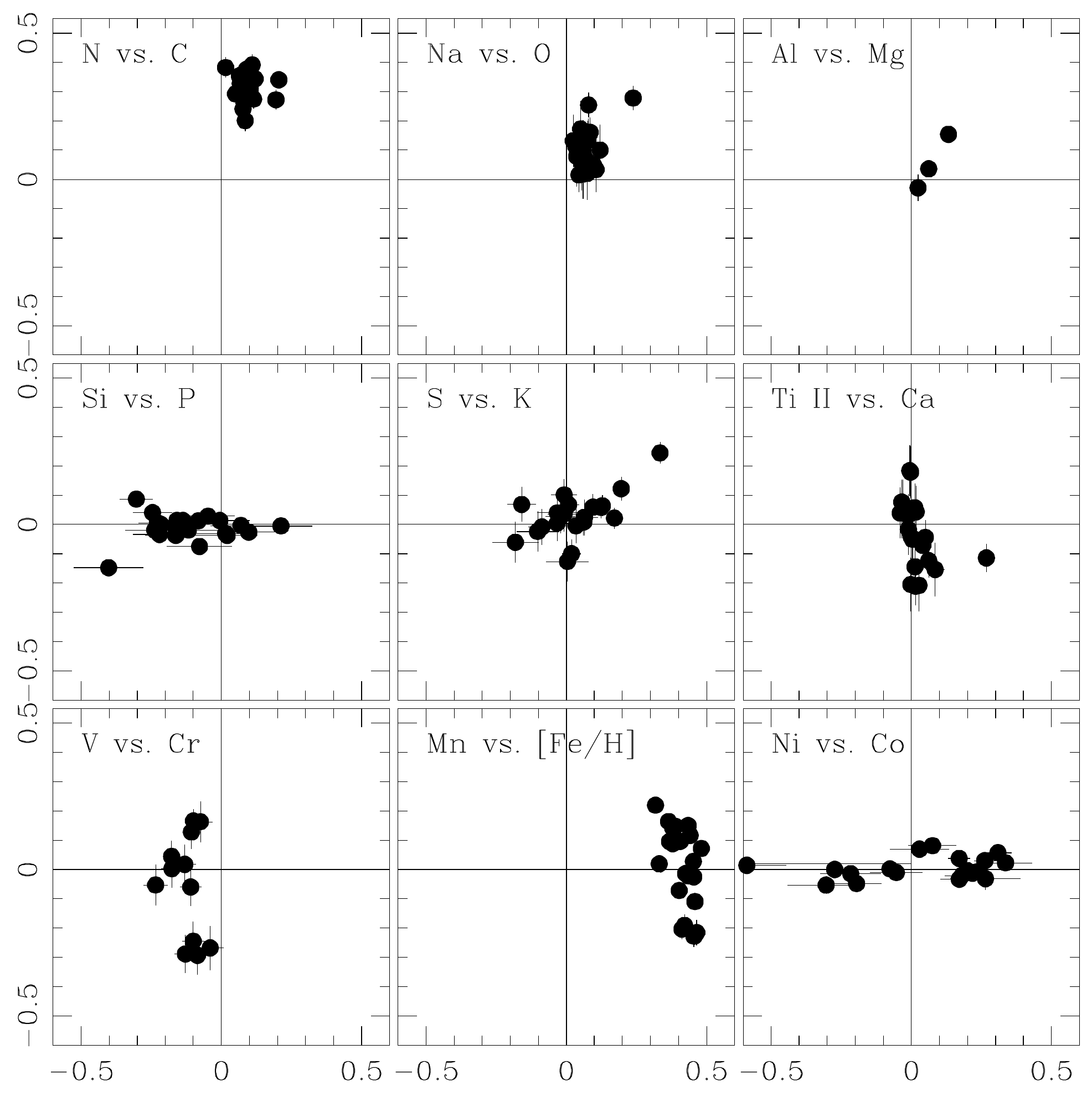}
       \includegraphics[width=0.9\columnwidth]{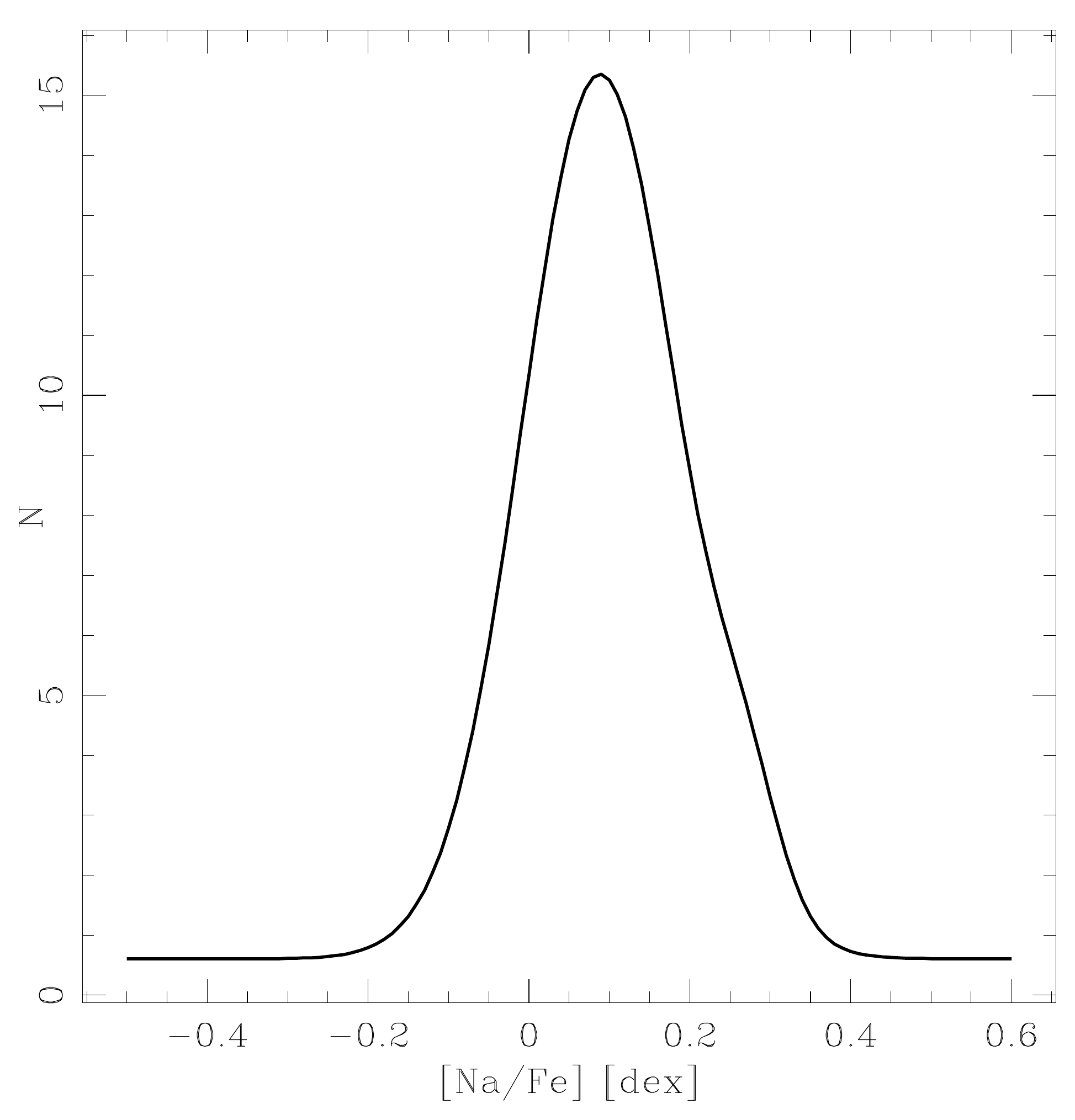}
\caption{Top, middle and bottom panels are the same as Figs.~\ref{fig1}-\ref{fig3},
for NGC\,6791, respectively.}
 \label{fig4}
\end{figure}

As a comparison, I consider NGC\,6791, which is also an old OC 
\citep[age = 7.7$\pm$0.4 Gyr, mass = 10.2$\pm$1.4 $\msun$,][]{bonatto2019} 
targered by \citet{bragagliaetal2014,cunhaetal2015} and
\citet{villanovaetal2018} as an MP candidate. None of these three studies found
evidence of MPs from the spectroscopic analysis of the observed cluster red
giants.  I took advantage of  the APOGEE DR14 database to search for abundances
of chemical elements of stars located in the cluster region. I found 21 stars
that satisfy the requirement of having RVs and [Fe/H] metallicities around the
OC's  values, namely, $<$RV$>$ = -47.37 $^{+011}_{-0.12}$ km/s
\citep{kamannetal2019} and $<$[Fe/H]$>$ = 0.40 $\pm$ 0.07 dex \citep{hetal14}.
I performed the same analysis carried out for NGC\,188 and found no evidence of
an  spread in [Na/Fe] (see Fig.~\ref{fig4}), in
full agreement with past studies. Therefore, the observed large Na spread in
NGC\,188 would not seem to be caused by a general abundance determination problem in
ASPCAP.

One of the causes of Na variations could in principle be non-LTE
effects. However, according to   \citet{cunhaetal2015}, any non-LTE {\em internal} Na variation among
the stars in our sample should remain well below 0.1~dex, and only a global
offset should apply. This would be true for both NGC\,188 and NGC\,6791.

\begin{figure}
     \includegraphics[width=\columnwidth]{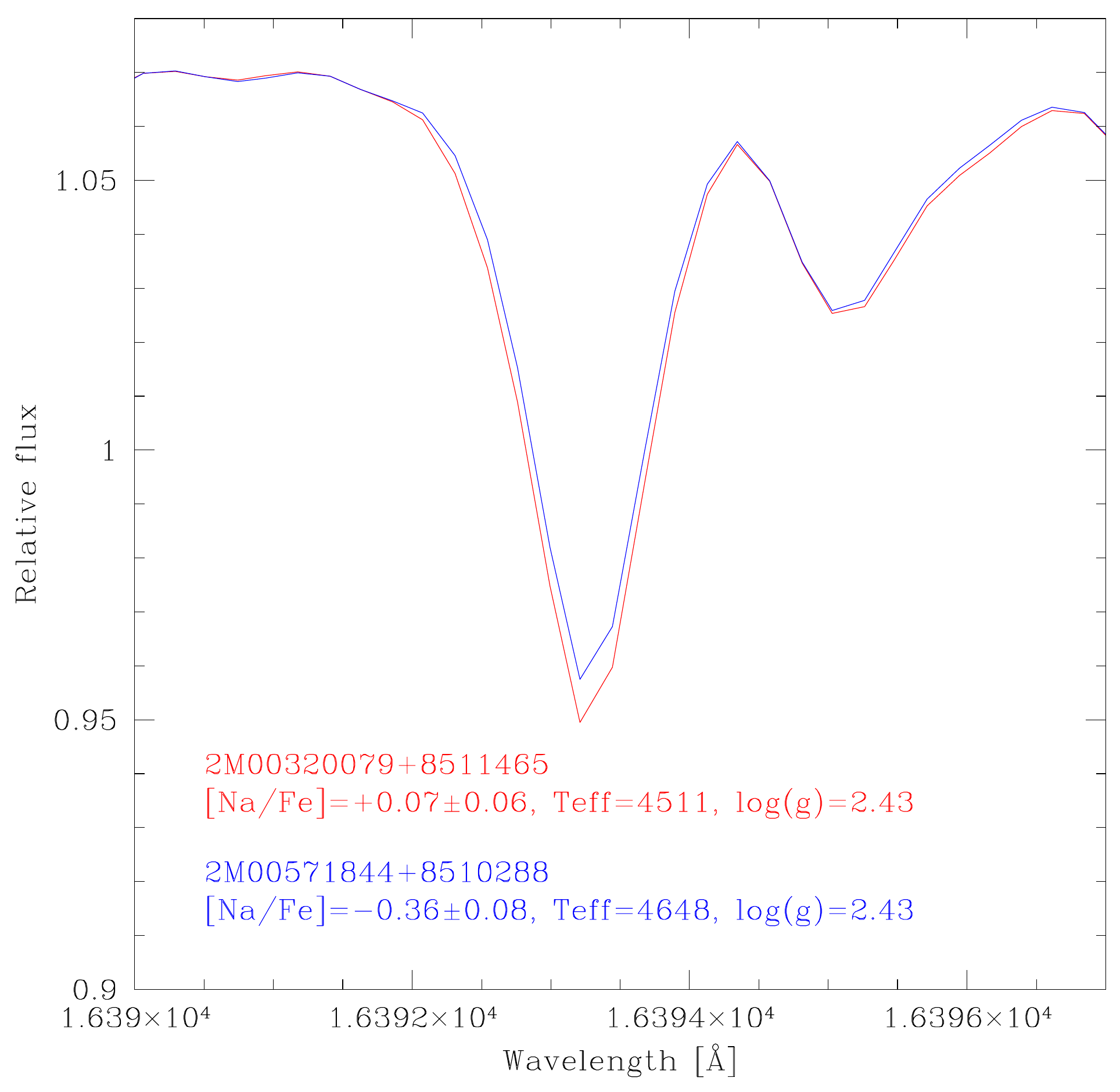}
\caption{Comparison of the spectra of two stars with similar atmospheric parameters,
respectively.}
 \label{fig5}
\end{figure}

However, the [Na/Fe] abundances have been derived from two
weak (in GK-giants) and possibly blended lines at about 16378 and 16393 \AA, respectively, 
for which \citet{jonssonetal2018} and 
\citet{holtzmanetal2018} showed that the Na uncertainty is of 0.16 dex.
Furthermore, I directly compared the available spectra for stars with similar atmospheric
parameters and it would seem that they look similar (see an example in Fig.~\ref{fig5}).
Therefore, I warn users of large spectroscopic surveys to be 
extra careful when finding peculiar abundance results. 
Analysis pipelines that run in an unsupervised fashion may produce bad results which may 
not be noticed before publication.


\begin{deluxetable}{lcc}
\tablecaption{Mean [X/Fe] ratios and intrinsic spreads. \label{tab:table2}}
\tablehead{
\colhead{Chemical} & \colhead{$<$[X/Fe]$>$} & \colhead{W}\\
\colhead{element} & \colhead{(dex)} & \colhead{(dex)} 
}
\startdata
C   & 0.00 $\pm$ 0.01  & 0.03  $\pm$ 0.01 \\
N   & 0.34 $\pm$ 0.01   & 0.03  $\pm$ 0.01 \\
O   &  0.01 $\pm$ 0.01  & 0.00  $\pm$ 0.01 \\
Na  & -0.04 $\pm$ 0.01  & 0.16  $\pm$  0.04 \\
Mg  & 0.04 $\pm$ 0.01  & 0.00  $\pm$ 0.01 \\
Al   & 0.06 $\pm$ 0.01  & 0.05 $\pm$  0.01 \\
Si   & 0.01 $\pm$ 0.01  & 0.02 $\pm$  0.01 \\
P    & -0.05 $\pm$ 0.02 & 0.00  $\pm$ 0.04 \\
S    & -0.01 $\pm$ 0.01  & 0.00  $\pm$ 0.01 \\
K    & 0.02 $\pm$ 0.01  & 0.02  $\pm$ 0.01 \\
Ca  & -0.02 $\pm$ 0.01 & 0.00 $\pm$  0.01 \\
Ti\,II &-0.07 $\pm$ 0.03  & 0.10  $\pm$ 0.03 \\
V    & 0.06 $\pm$ 0.02   & 0.08  $\pm$ 0.02 \\
V*    & 0.06 $\pm$ 0.02   & 0.05  $\pm$ 0.02 \\
Cr  & 0.01 $\pm$ 0.01  & 0.03  $\pm$ 0.01 \\
Mn   & 0.08 $\pm$ 0.01 & 0.02  $\pm$ 0.01 \\
Co      &  0.10 $\pm$ 0.02 &  0.05  $\pm$ 0.02 \\
Ni  &  0.03 $\pm$ 0.01 & 0.02  $\pm$ 0.01 \\
\enddata

\noindent (*) V  values after removal of
one very visible outlier in Fig.~\ref{fig2}.
\end{deluxetable}

\section{Conclusions}

I exploited the APOGEE DR14 database searching for stellar parameters and
element abundances of a wide variety of chemical species of stars in the field
of the old OC NGC\,188. I aimed at probing the existence of MPs as showed by
their different [X/Fe] ratios. MPs are commonly observed in Galactic GCs and in younger clusters down to $\sim$ 2 Gyr old and perhaps younger.
However, no galactic OC was confirmed to belong to the family of clusters with
MPs so far.

A crucial point for embarking in such a search relies on the availability of
accurate chemistry for a reasonable large sample of cluster members. While the
accuracy of the [X/Fe] ratios are tied to that of ASPCAP, cluster membership can
be assigned on the basis of stellar radial velocities, overall
metallicities, {\em Gaia} astrometry, and position with respect
to the cluster center. With these criteria, I could select a
sample of   15 {\it bonafide} cluster red giants.

From a simple maximum likelihood statistics of the [X/Fe] ratios for 12 
light-elements and 5 iron-peak ones, I found that those for Na, Al, Ti\,II, V,
and Co show dispersions larger than 0.05 dex, and up to 0.16 dex; the remaining
ones resulting with nearly null or null values. I noted that [X/Fe] differences
between first and second generation stars in GCs also exhibit
pattern of null values for some light-elements, while for others a significant
difference is observed; Na and N being the most remarkable
species that vary significantly in all GCs adequately observed so far.

By building the [X/Fe] distributions of the above mentioned 5 chemical elements,
I found that the derived spread in [Na/Fe] is caused by two groups of
measurements with a mean difference $\Delta$[Na/Fe] $\sim$ 0.30 dex. 
The absence of
a significant spread in N and a possible anti-correlation with Al -- while
generally Na and Al correlate in GCs -- would need to be fully understood. On
the other hand, the resulting distributions for V and Co showed that the derived
dispersions come from the presence of one isolated star which widen the whole
[X/Fe] range, and thus are not significant. Similarly, the fact that the large
Ti~II spread is not mirrored by a similar spread in Ti~I casts doubts on its
significance. Finally, the very similar shapes around the Na lines of the spectra 
of two stars with similar atmospheric parameters warned me to conclude on any 
intrinsic spread, but rather to be careful in its interpretation.

\begin{acknowledgements}
I am deeply indebted to E. Pancino
for her contribution to an early version of this work, who in turn
is warmly thankful to P. B. Stetson for providing his photometry for
NGC\,188 and for cross-matching it with {\em Gaia}. Funding for the Sloan
Digital Sky Survey IV has been provided by the Alfred P. Sloan Foundation,  the
U.S. Department of Energy Office of Science, and the Participating Institutions.
SDSS-IV  acknowledges support and resources from the Center for High-Performance
Computing at the University of Utah. The SDSS web site is www.sdss.org.

SDSS-IV is managed by the Astrophysical Research Consortium for the 
Participating Institutions of the SDSS Collaboration including the 
Brazilian Participation Group, the Carnegie Institution for Science, 
Carnegie Mellon University, the Chilean Participation Group, the French Participation Group, Harvard-Smithsonian Center for Astrophysics, 
Instituto de Astrof\'isica de Canarias, The Johns Hopkins University, Kavli Institute for the Physics and Mathematics of the Universe (IPMU) / 
University of Tokyo, the Korean Participation Group, Lawrence Berkeley National Laboratory, 
Leibniz Institut f\"ur Astrophysik Potsdam (AIP),  
Max-Planck-Institut f\"ur Astronomie (MPIA Heidelberg), 
Max-Planck-Institut f\"ur Astrophysik (MPA Garching), 
Max-Planck-Institut f\"ur Extraterrestrische Physik (MPE), 
National Astronomical Observatories of China, New Mexico State University, 
New York University, University of Notre Dame, 
Observat\'ario Nacional / MCTI, The Ohio State University, 
Pennsylvania State University, Shanghai Astronomical Observatory, 
United Kingdom Participation Group,
Universidad Nacional Aut\'onoma de M\'exico, University of Arizona, 
University of Colorado Boulder, University of Oxford, University of Portsmouth, 
University of Utah, University of Virginia, University of Washington, University of Wisconsin, 
Vanderbilt University, and Yale University.

This work presents results from the European Space Agency (ESA)
space mission Gaia. Gaia data are being processed by the Gaia Data Processing
and Analysis Consortium (DPAC). Funding for the DPAC is provided by national
institutions, in particular the institutions participating in the Gaia
MultiLateral Agreement (MLA). The Gaia mission website is
\url{https://www.cosmos.esa.int/gaia}. The Gaia archive website is
\url{https://archives.esac.esa.int/gaia}. 
\end{acknowledgements}





\end{document}